\documentclass[12pt]{article}
\usepackage{amsmath,amssymb,latexsym,graphicx,psfrag,colortbl,booktabs,cite}
\newcommand{\gr}{\rowcolor[gray]{.9}}

\begin{document}

\begin{titlepage}

\begin{flushright}
FERMILAB-PUB-07-406-T\\
SI-HEP-2007-11\\[0.2cm]
August 6, 2007
\end{flushright}

\vspace{0.7cm}
\begin{center}
\Large\bf
\boldmath
Kinetic corrections to $\bar B\to X_c \ell \bar \nu$ at one loop
\unboldmath
\end{center}

\vspace{0.8cm}
\begin{center}
{\sc Thomas Becher$^a$, Heike Boos$^b$ and Enrico Lunghi$^a$}\\
\vspace{0.4cm}
{\sl $^a$\,Fermi National Accelerator Laboratory\\
P.O. Box 500, Batavia, IL 60510, U.S.A.\\[0.3cm]
$^b$ Theoretische Physik 1, Fachbereich Physik, Universit\"at Siegen\\
D-57068 Siegen, Germany
}
\end{center}

\vspace{1.0cm}
\begin{abstract}
\vspace{0.2cm}
\noindent 
We evaluate the one-loop corrections to the Wilson coefficient of the kinetic operator in the operator product expansion of the differential $\bar B\to X_c \ell\bar\nu$ decay rate. With a moderate cut on the lepton energy, the one-loop terms change the kinetic operator contributions to spectral moments by about $20\%$.  This amounts to a small correction for  leptonic and hadronic moments, except for those which vanish at the lowest order, where the effect can be sizable. Together with a two-loop calculation of the leading-power rate and an evaluation of the one-loop corrections to the Wilson coefficient of the chromo-magnetic operator, our results will allow for a high-precision determination of $|V_{cb}|$ and the $b$- and $c$-quark masses.
\end{abstract}
\vfil

\end{titlepage}

\section{Introduction}

Inclusive $\bar B\rightarrow X_c\ell\bar\nu$ decays are a precise probe of the underlying $b$- to $c$-quark transition because hadronisation effects are small and have a simple structure. These effects are suppressed by powers of the heavy-quark mass and given in terms of a small number of non-perturbative parameters. In the heavy-quark limit, the hadronic decay rate becomes equal to the partonic decay rate. The leading corrections are of order $1/m_b^2$ and are given in terms of two non-perturbative heavy-quark parameters, $\mu_\pi^2$ and $\mu_G^2$ which are the $B$-meson matrix elements of the kinetic and chromo-magnetic operator respectively. Schematically, the decay rate takes the form \cite{Chay:1990da, Bigi:1993fe,Blok:1993va,Manohar:1993qn}
\begin{equation}\label{eq:rate}
\Gamma(\bar B\rightarrow X_c \ell \bar\nu) = \frac{G_F |V_{cb}|^2 m_b^5}{192\pi^3} \left\{f(\rho) + k(\rho) \frac{\mu_\pi^2}{2m_b^2} + g(\rho) \frac{\mu_G^2}{2m_b^2}+ {\cal O}(m_b^{-3})
\right\} \, ,
\end{equation}
where $\rho=m_c^2/m_b^2$. The coefficients $f$, $g$ and $k$ can be calculated in perturbation theory:
\begin{equation}
f=f^{(0)}(\rho) + \frac{\alpha_s}{\pi}\, f^{(1)}(\rho) + \left(\frac{\alpha_s}{\pi}\right)^2\, f^{(2)}(\rho) + {\cal O}(\alpha_s^3) \,, \mbox{etc. }
\end{equation}
The general structure of the expansion is the same for other observables, such as partial rates or moments of the decay spectrum,  but the calculable coefficients $f$, $g$ and $k$ are different. For the total rate the kinetic corrections have the same coefficient as the leading order, $k(\rho)=-f(\rho)$. Also for other observables, such as partial rates and moments, the kinetic corrections can be obtained from the leading-power differential rate, but the relations are more complicated and are evaluated to ${\cal O}(\alpha_s)$ for the first time in this paper.

To turn (\ref{eq:rate}) into a precision determination of $|V_{cb}|$ one needs the values of $m_b$, $m_c$ and the heavy-quark parameters. Since the same parameters enter moments of the decay spectrum, one can determine these parameters by measuring not only the rate, but also a number of  moments. To this end, lepton energy moments and hadronic invariant mass and energy moments are measured \cite{Aubert:2004td, Aubert:2004te, Csorna:2004kp, Mahmood:2004kq, Abdallah:2005cx, Acosta:2005qh, Urquijo:2006wd, Schwanda:2006nf}. Using the results of these measurements, several groups have performed fits of the theoretical expressions to the experimental data \cite{Bauer:2002sh, Bauer:2004ve, Buchmuller:2005zv,Abe:2006xq}. The theoretical expressions that are used in the fit include one-loop corrections to the leading-power coefficients $f(\rho)$ \cite{Jezabek:1988iv, Jezabek:1988ja, Czarnecki:1989bz, Czarnecki:1994pu, Voloshin:1994cy, Falk:1995me, Gambino:2004qm, Trott:2004xc,Uraltsev:2004in, Aquila:2005hq} as well as the $\beta_0\alpha_s^2$-part of the two-loop corrections \cite{Luke:1994du,Ball:1995wa,Gremm:1996gg, Falk:1997jq, Aquila:2005hq}, while the coefficients $g(\rho)$ and $k(\rho)$ of the power corrections are known only at the tree level. In addition to the second-order power corrections proportional to $\mu_\pi^2$ and $\mu_G^2$, the fits also include the third-order power corrections, which involve two additional hadronic parameters, $\rho_D^3$ and $\rho_{LS}^3$ \cite{Gremm:1996df} (the fourth order corrections are now available as well \cite{Dassinger:2006md}). This technique yields the most precise determination of $|V_{cb}|$ together with very precise determinations of the heavy-quark masses. Already now, the estimated theoretical uncertainties are somewhat larger than the experimental ones \cite{Buchmuller:2005zv}. In the future the experimental uncertainty will decrease further: the BaBar moment measurements which were used in \cite{Bauer:2004ve, Buchmuller:2005zv} were published in 2004 and are based on $50\,{\rm fb}^{-1}$ of data \cite{Aubert:2004td,Aubert:2004te}\footnote{Very recently, Babar has presented preliminary results for hadronic moments based on $210 \,{\rm fb}^{-1}$ \cite{Collaboration:2007ya}.}, and the recently published Belle measurements on $140\, {\rm fb}^{-1}$ \cite{Urquijo:2006wd,Schwanda:2006nf}, but combined the two experiments have already collected more than $1\,{\rm ab}^{-1}$ of data. Also, based on the convergence of the perturbative series of the rate for $\tau$-decay and based on the size of the two-loop contributions that arise when converting the theoretical expressions between different schemes used in the literature, it has been suggested that the theoretical uncertainties in the results of the moment fits might be underestimated \cite{neubertFPCP07}. Whether the uncertainties are reliable is an important question because the value of $m_b$ extracted from the fit is a crucial ingredient for the determination of $|V_{ub}|$ from inclusive decays. After imposing the severe cuts necessary to eliminate the charm background, the prediction for the $\bar B\rightarrow X_u\ell\bar \nu$ rate behaves as $m_b^{n}$ with $n\approx 10-15$ \cite{Neubert:2001ib,Lange:2005yw}. The value and uncertainty of the extracted $|V_{ub}|$ are thus strongly correlated with the value and uncertainty of $m_b$.

It is clearly desirable to increase the precision of the theoretical predictions. To achieve this goal, two ingredients are needed: the leading-power moments have to be evaluated to two-loop accuracy, and the coefficients of the power corrections proportional to $\mu_\pi^2$ and $\mu_G^2$ need to be evaluated to one loop. In this paper we take the first and simplest step in this direction by evaluating the coefficient of the kinetic operator to one-loop accuracy. Let us stress that, while it is demanding, also the ${\cal O}(\alpha_s^2)$ calculation of the leading-power moments is doable. A few years ago such a calculation looked prohibitively difficult, but in the meantime the necessary methods to perform it numerically have been developed \cite{Binoth:2000ps, method}. Indeed, the two-loop correction for muon decay  $\mu \to X_e \nu \bar\nu$, the QED equivalent of $\bar B\to X_c \ell\bar\nu$, has been evaluated recently using this method \cite{Anastasiou:2005pn}. We use the same numerical approach for our one-loop calculation, because the size of the expressions involved is such that an analytic calculation does not look feasible. Since the method is gauge invariant (we perform the calculation without introducing a gluon mass), it is also suited for the calculation of the corrections to the coefficient of the chromomagnetic operator.

The kinetic corrections are obtained by expanding the leading-power ${\cal O}(\alpha_s)$ expressions up to second order in the small residual momentum of the $b$-quark inside the $B$-meson. To have a check of our results, we perform the calculation in two different ways. A straightforward and tedious way of obtaining the kinetic corrections is to expand the leading-power Feynman diagrams in the residual momentum before performing the loop and phase-space integrations. The resulting expressions are long and involve terms which are individually strongly infrared divergent. Another complication arises because the expansion produces not only standard phase-space integrals, but also derivatives of such phase-space integrals, which arise from cutting higher powers of propagators. A much more elegant and efficient way to perform the calculation is to expand the result for the leading-order differential rate in the residual momentum. In this way one obtains results for the moments in terms of integrals over the leading-power rate and its derivative. In fact, without experimental cuts one obtains simple algebraic relations between the kinetic corrections and the leading-power moments. As a byproduct of our analysis, we obtain the ${\cal O}(\alpha_s)$ corrections to the leading-power moments and reproduce the numerical results of \cite{Aquila:2005hq}. 

As the two-loop corrections to the leading-power rate and the one-loop corrections to the $\mu_G^2$ terms are not yet available, it is too early to perform a detailed phenomenological analysis. Instead, we present numerical values for a few reference values of the electron energy cut. For the moments which do not vanish, the corrections we calculate turn out to be small, below $1\%$ as long as the cut on the lepton energy is not too strong. For the moments of the partonic invariant mass $(p_x^2-m_c^2)^n$, on the other hand, which vanish at the tree-level and leading power, the corrections are larger, of order $30\%$. We expect the corrections proportional to $\mu_G^2$ to be more important than the kinetic corrections. For the tree-level rate, they are roughly a factor ten larger than the kinetic corrections.

This paper is organized as follows. In the next section, we explain how the kinetic corrections are calculated using the operator product expansion (OPE). In Section \ref{phasespace} we give parameterizations for the phase-space and loop integrals which are needed to perform the calculation. These parameterizations map the integration onto the unit hypercube and are such that infrared divergences appear only in a single variable and can be isolated before performing the numerical integration. We present our numerical results in Section \ref{results}. In the same section, we give formulae for the conversion of partonic to hadronic moments. We also sketch discuss how to convert our results into different schemes, such as the kinetic \cite{Uraltsev:1996rd}, potential-subtracted \cite{Beneke:1998rk}, $1S$ \cite{Hoang:1998hm} or the shape-function scheme \cite{Bosch:2004th}.

\section{Evaluation of the kinetic corrections\label{kinetic}}
In this section we briefly explain how we evaluate the $\bar B\rightarrow X_c\ell\bar\nu$ decay rate using the OPE. The application of the OPE to inclusive $B$ decays has been worked out quite some time ago and the reader interested in more details should consult the original references \cite{Chay:1990da, Bigi:1993fe,Blok:1993va,Manohar:1993qn} or the textbook \cite{Manohar:2000dt}. Our goal in this section is to outline the necessary steps to obtain the result and to discuss some of the technicalities which are encountered in the course of the one-loop calculation. The reader solely interested in the numerical results can skip to section \ref{results}. 

The $\bar B\rightarrow X_c\ell\bar\nu$ decay is mediated by the effective Hamiltonian
\begin{equation}
{\cal H}_{\rm eff}=\frac{G_F}{\sqrt{2}}V_{cb}\,\, {J}^\mu\, J^{\ell}_\mu=  \frac{G_F}{\sqrt{2}}V_{cb}\, {\bar c}\,\gamma^\mu\,(1-\gamma_5)\,b \; \bar\ell\,\gamma_\mu\,(1-\gamma_5)\,\nu \,.
\end{equation}
The decay rate factors into a leptonic tensor $L_{\mu\nu}$ and a hadronic tensor $W_{\mu\nu}$
\begin{equation}
{\rm d}\Gamma=\frac{G_F^2\,|V_{cb}|^2}{2}\,\,  {\rm d}\mu(p_\ell) \, {\rm d}\mu(p_\nu) \, L_{\mu\nu}(p_\ell,p_\nu) W^{\mu\nu}(p_B,q)\, ,
\end{equation}
where $q=p_\ell+p_\nu$ and ${\rm d}\mu(p)$ denotes the phase space
\begin{equation}
{\rm d}\mu(p)= \frac{{\rm d}^{d-1} p}{(2\pi)^{d-1}\, 2E}\,
\end{equation}
in $d=4-2\epsilon$ dimensions. Since the differential rate is a finite quantity, we could set $d=4$. However, individual contributions to the hadronic tensor contain ultra-violet (UV) as well as infrared (IR) divergences which we regulate by keeping $\epsilon\neq 0$ throughout. The spin-averaged leptonic tensor is
\begin{equation}
L_{\mu\nu}= {\rm Tr}\left[p\!\!\!/_\ell\, \gamma_\mu(1-\gamma_5)\,p\!\!\!/_\nu  \gamma_\nu(1-\gamma_5) \right]\, .
\end{equation}
The hadronic tensor is obtained by taking the imaginary part of the 
time-ordered products of currents
\begin{equation}
W_{\mu\nu}=-2\,{\rm Im}\, T_{\mu\nu},
\end{equation}
where
\begin{equation}\label{Tmunu}
T_{\mu\nu}= -i\int\!d^4q e^{-iq x} \frac{1}{2M_B} \langle \bar B(p_B)|\,{\bf T}\left[ J_\mu^\dagger(x)\, J_\nu (0)\right] |\bar B(p_B)\rangle\,.
\end{equation}
We work in the kinematics $p_B^\mu=M_B v^\mu$ and our states are canonically normalized. In analytic calculations the hadronic tensor is usually decomposed into five form factors, but we prefer to directly evaluate the relevant product $W_{\mu\nu} L^{\mu\nu}$. Since we use dimensional regularization we need to specify how we treat $\gamma_5$ in $d$ dimensions. A definition of the axial current in $d$ dimensions suitable for our purposes has been given by Larin \cite{Larin:1993tq} and we adopt it for our calculation.

The operator product appearing in (\ref{Tmunu}) is expanded in a series of local operators which corresponds to an expansion of the rate in inverse powers of the $b$-quark mass. To perform the expansion, one first removes a rapidly oscillating factor from the $b$-quark field by writing it as $b(x)=e^{-imvx} b_v(x)$. Using heavy-quark effective theory (HQET) \cite{Neubert:1993mb,Manohar:2000dt}, all the matrix elements necessary to second order in the expansion can be reduced to
\begin{align}
\langle O_3 \rangle &\equiv \frac{1}{2M_B}\langle \bar B(p_B)|\, \bar b_v\, v\!\!\!/\, b_v\, | \bar B(p_B)\rangle = 1 \,, \nonumber \\
\langle O_{\rm kin} \rangle&\equiv \frac{1}{2M_B} \langle \bar B(p_B)|\, \bar{b}_v (i D)^2 {b}_v\, |\bar B(p_B)\rangle = - \mu_{\pi}^2\, , \label{matrixelements} \\
\langle O_{\rm mag}\rangle &\equiv \frac{1}{2M_B}\langle \bar B(p_B)|\, \bar{b}_v \frac g2 \sigma_{\mu\nu}G^{\mu\nu} {b}_v\, |\bar B(p_B)\rangle = \mu_{G}^2 \,.\nonumber
\end{align}
The HQET parameter $\mu_\pi^2$ is often denoted by $-\lambda_1$ and is not renormalized. Up to terms suppressed by three powers of the heavy-quark mass, the decay rate thus takes the form
\begin{equation}
{\rm d}\Gamma = \frac{G_F^2\,|V_{cb}|^2}{2}\,\,  {\rm d}\mu(p_\ell) \, {\rm d}\mu(p_\nu) \left [ C_3(v,p_\ell,p_\nu) \langle O_3 \rangle + C_{\rm kin}(v,p_\ell,p_\nu) \langle O_{\rm kin} \rangle + C_{\rm mag}(v,p_\ell,p_\nu) \langle O_{\rm mag} \rangle   \right] \, .\nonumber
\end{equation}
The Wilson coefficients $C_i(v,p_\ell,p_\nu)$ of the three operators are independent of the external states and can be calculated using partonic initial and final states. To extract the coefficient $C_{\rm kin}(v,p_\ell,p_\nu)$ of the kinetic operator, it is simplest to use an on-shell $b$-quark with momentum $p_b=m_b v_\mu+r_\mu$, which amounts to calculating the partonic decay rate $b\rightarrow X_c \ell \bar\nu$. To find the coefficient of the operator $O_{\rm kin}$ with two derivatives, we expand the partonic rate to second order in the residual momentum $r_\mu$. The result takes the form
\begin{equation}
{\rm d}\Gamma^{\rm partonic} = A + A_\mu\, \frac{1}{m_b}\,r^\mu + A_{\mu\nu}\,\frac{1}{m_b^2}\,r^\mu r^\nu +{\cal O}(r^3)\,.
\end{equation}
At the loop level, the question arises whether to expand the diagrams before or after the loop integration. In general, either choice is valid as long as one evaluates the loop corrections to the operator product and to the matrix elements of the local operators $O_i$ in the same way. In our case the situation is especially simple: since we perform the matching calculation on-shell, the one-loop corrections to the HQET matrix elements of the operators $O_i$  vanish and the loop integration commutes with the expansion in the residual momentum. To have a check of our results, we evaluate the corrections in both ways.

We can further simplify the calculation by averaging over the direction of the transverse momentum $r_\perp^\mu=r^\mu-v\cdot r\, v^\mu$. The component parallel to $v_\mu$ is fixed by the on-shell condition $2m_b v\cdot r=-r^2$. Taking the average we have
\begin{equation}
{\rm d}\Gamma^{\rm partonic} = A - A_\mu\, v^\mu \frac{r^2}{2m_b^2} +A_{\mu\nu}\, \frac{r^2}{m_b^2}\, \frac{1}{d-1} (g_{\mu\nu}-v_\mu v_\nu) +{\cal O}(r^3)\, .
\end{equation}
To obtain the hadronic rate, we first bring the leading-power partonic matrix element into the form (\ref{matrixelements}) by rewriting
\begin{align} \label{eq:matrix}
\langle  b(p_b)|\, {\bar b}_v\,  \,b_v\, |  b(p_b)\rangle & = 
\frac{1}{m_b} \langle b(p_b)|\, {\bar b}_v\, {p\!\!\!/}_b\, b_v\, |  b(p_b)\rangle
\nonumber \\
&=\langle  b(p_b)|\, {\bar b}_v\, {v\!\!\!/}\, b_v\, |  b(p_b)\rangle
+\frac{r^2}{2m_b^2} \langle  b(p_b)|\, {\bar b}_v\, b_v\, | b(p_b)\rangle\,.
\end{align}
We then replace the partonic matrix elements by the corresponding hadronic matrix elements:
\begin{align}\label{eq:hadronic}
{\rm d}\Gamma =& A\, \frac{1}{2M_B}\langle \bar B(p_B)|\, {\bar b}_v\, v\!\!\!/\, b_v\, | \bar B(p_B)\rangle \nonumber\\
&+ \left[ A -A_\mu\, v^\mu + A_{\mu\nu} \frac{2}{d-1} (g_{\mu\nu}-v_\mu v_\nu) \right] \,\frac{1}{2M_B} \langle \bar B(p_B)|\, {\bar b}_v\, (iD)^2 b_v\, | \bar B(p_B)\rangle +\dots \nonumber \\
=& A-\frac{\mu_\pi^2}{2m_b^2}\left[A-A_\mu\, v^\mu + A_{\mu\nu} \frac{2}{d-1} (g_{\mu\nu}-v_\mu v_\nu)\right] +\dots\,.
\end{align}
The ellipsis denotes terms which are suppressed by $m_b^{-3}$ or proportional to $\mu_G^2$. The operator $O_{\rm mag}$ has a vanishing $b$-quark matrix element and its coefficient is therefore not determined by our matching calculation. 

\begin{figure}[t]
\begin{center}
\psfrag{b}[B]{$m_b v+r$}
\psfrag{q}[B]{$q$}
\psfrag{c}[B]{$m_b v+r-q$}
\includegraphics[width=0.35\textwidth]{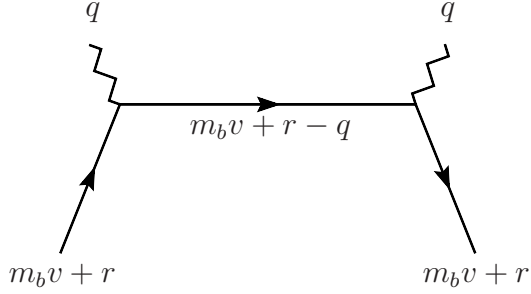}
\end{center}
\caption{Tree-level contribution to the hadronic tensor.
\label{WmunuTree}}
\end{figure}

To illustrate the structure of the result, we now calculate the kinetic corrections at tree level. The tree-level contribution to the hadronic tensor is shown in Figure 
\ref{WmunuTree}. Taking its imaginary part and contracting with the leptonic tensor, the partonic tree-level rate is found to be
\begin{equation}
{\rm d}\Gamma^{\rm partonic}=32\, G_F^2\,|V_{cb}|^2\,\,  {\rm d}\mu(p_\ell) \, {\rm d}\mu(p_\nu) \, \, p_b\cdot p_\nu \,(p_b-q) \cdot p_\ell \, (2\pi)\,\delta((p_b-q)^2-m_c^2)\, .
\end{equation}
We now expand up to second order in the residual momentum $r_\mu$, average over the $\perp$-direction and replace the partonic by the hadronic matrix elements to obtain the result
\begin{multline}
{\rm d}\Gamma = 64 \pi \, G_F^2\,|V_{cb}|^2\,\,  {\rm d}\mu(p_\ell) \, {\rm d}\mu(p_\nu) \, \Bigg[ f_0 \,\delta \left(p_c^2-m_c^2\right)+f_1\,\delta '\left(p_c^2-m_c^2\right) 
+f_2\,\delta'' \left(p_c^2-m_c^2\right)\Bigg]\,,
\end{multline}
where $p_c=m_b v-q$ and
\begin{align*}
f_0 &=m_b v\cdot p_\nu p_c\cdot p_\ell +\frac{\mu_\pi^2}{2m_b^2} \frac{m_b}{3} \Big [ 5 m_b  v\cdot 
p_\ell v\cdot p_\nu -2 m_b\, p_\ell\cdot p_\nu \Big] \,,\\ 
f_1&=\frac{\mu_\pi^2}{3} \Big[ v\cdot p_c  v\cdot p_\nu\, (2 
m_b v+5 p_c)\cdot p_\ell - p_c\cdot p_\ell \,(5 m_b v+2 p_c)\cdot p_\nu \Big] \,, \\
f_2&=2\,\mu_\pi^2\, m_b\,
 v\cdot p_\nu p_c\cdot p_\ell\, \left[(v\cdot p_c)^2- p_c^2 \right]\,.
  \end{align*}
 The one-loop results for the rate have a similar structure, also in this case the result contains $\delta^{(n)}\left(p_c^2-m_c^2\right)$ with $n=0,1,2$. To calculate moments of the decay spectrum, we introduce an integration over the partonic phase space
\begin{equation}\label{partonicphase}
1=\int \frac{dp_x^2}{(2\pi)} \int {\rm d} \mu(p_x)\, (2\pi)^d\,\delta^d(m_b v-q-p_x) \, .
\end{equation}
The result for the decay rate then takes the form
\begin{equation}\label{treeratediff}
\Gamma= \left. \int \left [ {\rm d} \Pi\right] f_0\right|_{m_x^2=m_c^2}
-\left. \frac{\rm d}{{\rm d} m_x^2} \int \left [ {\rm d} \Pi \right] f_1\right|_{m_x^2=m_c^2}
+\left. \frac{\rm d^2}{{\rm d} (m_x^2)^2} \int \left [ {\rm d} \Pi \right] f_2\right|_{m_x^2=m_c^2}\, ,
\end{equation}
where we have used the notation
\begin{equation}\label{treephasespace}
\int \left [ {\rm d} \Pi\right] \equiv 
\int \left [ {\rm d} \Pi_{b\rightarrow x+\ell+\bar\nu}\right] = 
\int\! {\rm d}\mu(p_x) \int\! {\rm d}\mu(p_\ell) \int\! {\rm d}\mu(p_\nu) \, (2\pi)^d \delta^d(m_b v-p_x-p_\ell-p_\nu)\, .
\end{equation}
It turns out that for the total rate the derivative terms in (\ref{treeratediff})  do not contribute. However, for partial rates or spectral moments these terms do give non-vanishing contributions. To evaluate (\ref{treeratediff}) numerically, we need a suitable parameterization for the phase-space integral (\ref{treephasespace}). The necessary parameterizations, both for the  tree-level phase space and the phase space with the emission of an additional gluon, needed for the ${\cal O}(\alpha_s)$ corrections to the rate, are given in the next section.

An alternative, more elegant and efficient way of evaluating the kinetic corrections was described in \cite{Manohar:2000dt}. Instead of expanding the diagrams which contribute to the hadronic tensor, one takes the result for the differential partonic rate and expands in the residual momentum. To derive this result, it is convenient to introduce the dimensionless variables $x=2\,E_\nu/m_b$, $y=2\,E_\ell/m_b$ and $\hat{q}^2=q^2/m_b^2$. When expanding in the residual momentum, one has
\begin{align}
x &\rightarrow x + \frac{2}{m_b}\, r \cdot p_\nu\, &
y &\rightarrow y + \frac{2}{m_b}\, r \cdot p_\ell\, &
\hat q^2 &\rightarrow \hat q^2\,.
\end{align}
Expanding to second order and averaging over the $\perp$-direction, the hadronic differential rate is equal to \cite{Manohar:2000dt}
\begin{multline}
\frac{{\rm d}\Gamma}{ {\rm d}x\, {\rm d}y\,{\rm d}{\hat q}^2}
=  \left[ 1 + \frac{\mu_\pi^2}{2m_b^2}\left ( -1 +x \frac{\partial}{\partial x} +y \frac{\partial}{\partial y} +\frac{1}{3}\,x^2 \frac{\partial^2}{\partial x^2} +\frac{1}{3}\,y^2 \frac{\partial^2}{\partial y^2}  
\right.\right. \\ \left.\left.
+\frac{2}{3} (x y - 2 \hat{q}^2) \frac{\partial^2}{\partial x \partial y}\right)\right] \frac{{\rm d}\Gamma^{\rm partonic}}{ {\rm d}x\, {\rm d}y \,{\rm d}\hat{q}^2}\, .
\end{multline}
The $\mu_\pi^2$-term without derivatives comes from expanding the matrix elements, see (\ref{eq:matrix}). What makes this result particularly useful is that we can explicitly evaluate the derivatives using integration by parts when calculating moments. For the moments with a cut on the lepton energy
\begin{equation}
\left [ x^n\, y^m\, ({\hat q}^2)^l  \right]_{y_0} = \int \! {\rm d}x\,{\rm d}y \,{\rm d}\hat{q}^2 \, \frac{{\rm d}\Gamma}{ {\rm d}x\, {\rm d}y\,{\rm d}{\hat q}^2} \,x^n\, y^m\, ({\hat q}^2)^l \theta(y-y_0) \, ,
\end{equation}
one finds
\begin{multline} \label{eq:momrel}
\left [ x^n\, y^m\, ({\hat q}^2)^{l}   \right]_{y_0} =\left [ x^n\, y^m\, ({\hat q}^2)^{l}   \right]_{y_0}^{\rm partonic} \\
+ \frac{\mu_\pi^2}{6m_b^2} 
\left [  \left((n+m)^2+2m+2 n-3\right) x^n\,
   y^m\, ({\hat q}^2)^l-4\, m\, n \, x^{n-1}\, y^{m-1}\, ({\hat q}^2)^{l+1}  \right]_{y_0}^{\rm partonic} \\
   + \frac{\mu_\pi^2}{6m_b^2} \left[ \left((m+2n+1) x y_0 -4 n \hat{q}^2 \right) x^{n-1} y_0^{m} ({\hat q}^2)^{l} \, \delta(y-y_0) + x^{n} y_0^{m+2} ({\hat q}^2)^{l} \delta'(y-y_0) \right]^{\rm partonic} \,.
   \end{multline}
   The terms in the third line are boundary terms and vanish when setting $y_0=0$. In this case the kinetic corrections to the moments follow via simple algebraic relations from the leading-power term. The explicit relations for the moments we are interested in are given in Appendix \ref{app:momrel}. In the general case with a cut on the lepton energy $y_0\neq 0$, one needs to also evaluate moments of the partial rate and its first derivative. To evaluate these boundary terms, it is important to keep in mind that $\hat{q}^2$, $x$ and $y$ are not completely independent: the rate includes a factor $\theta(x y -{\hat q}^2)$ and at tree level the variables fulfill $1-\frac{m_c^2}{m_b^2}+{\hat q}^2=x+y$. More generally, when choosing a phase-space parameterization to evaluate (\ref{eq:momrel}), the variables $\hat q^2$ and $x$ become functions of $y$ and the derivative in the third line of (\ref{eq:momrel}) then acts not only on the rate but also on the factors $x^a  ({\hat q}^2)^{b}$.

\section{Phase-space and loop integrals\label{phasespace}}

\begin{figure}[t]
\begin{center}
\begin{tabular}{cccc}
\includegraphics[width=0.23\textwidth]{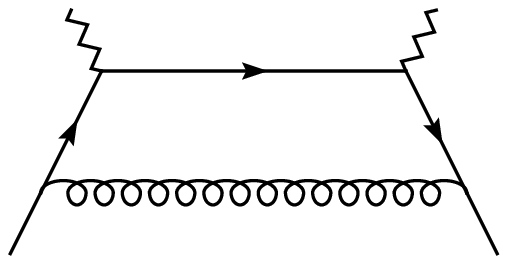} 
&  \includegraphics[width=0.23\textwidth]{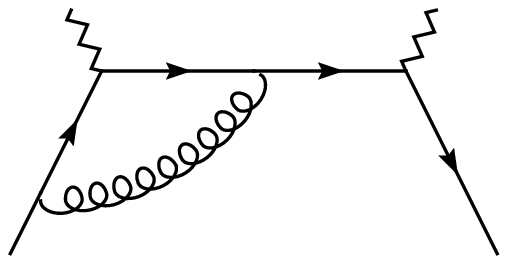} &
 \includegraphics[width=0.23\textwidth]{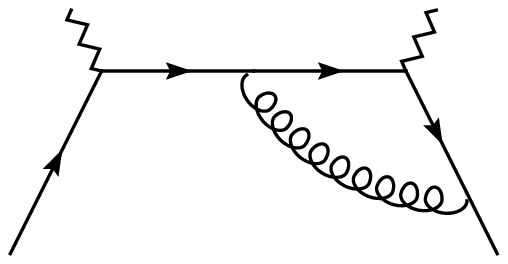} 
&  \includegraphics[width=0.23\textwidth]{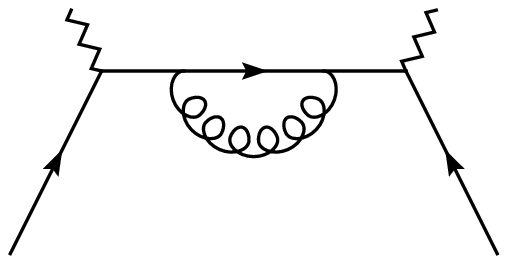} 
\end{tabular}
\end{center}
\caption{One-loop contributions to the hadronic tensor.\label{fig:loop}}
\end{figure}

We now derive phase-space representations which are well suited for the numerical calculation of the kinetic corrections at one loop. The diagrams contributing to the hadronic tensor are shown in Figure \ref{fig:loop}. Their imaginary part receives contributions from virtual corrections as well as real-emission contributions.  In the real-emission contributions, the imaginary part of the diagram is generated by an intermediate state with an on-shell gluon and on-shell charm quark, so we need a parameterization of the $b\rightarrow c+g+\ell+\bar\nu$ phase space. The loop integrations necessary to evaluate the virtual corrections contain UV as well as IR divergences. The real contributions are ultraviolet finite, but contain IR divergences which cancel against the IR divergences of the virtual corrections. Because the quarks are massive, soft gluons are the only source of infrared divergences at one loop.

To allow for a simple numerical evaluation, we map the phase-space and loop integrations to the unit hypercube. Also, since we want to calculate the rate and spectral moments with a cut on the lepton energy, we choose a parameterization in which the lepton energy is one of the variables. A last requirement is that we want the infrared divergences to be restricted to a single variable, so that they are easily isolated. It is convenient to split the phase-space integral into a hadronic and leptonic part
\begin{equation}
\int \left [ {\rm d} \Pi_{b\rightarrow  c + g+\ell +\bar \nu }\right] = \int \frac{d p_x^2}{2\pi} \int \left [ {\rm d} \Pi_{b\rightarrow x + \ell + \bar\nu}\right]  \int \left [ {\rm d} \Pi_{x\rightarrow c + g}\right] \,.
\end{equation}

\subsection{Three-body phase space $b\rightarrow x + \ell + \bar\nu$} 

We denote the phase-space integration variables by $\lambda_i \in [0,1]$ with $i=1\dots 4$. Neglecting the lepton masses, we choose the momenta as 
\begin{align}
{p_b}&=(m_b,0,0,0) \,,& p_\ell &= (E_\ell,0,0, E_\ell)  \,,&
p_\nu&=(E_\nu,E_\nu \sin\theta_1,0, E_\nu \cos\theta_1 )  \,,
\end{align}
and parameterize 
\begin{align}
E_\ell &= m_b\, \frac{y}{2},&
E_\nu &= m_b\,\frac{(1-\rho-y)\,(1-\lambda _2)}{2 \kappa} ,& \cos\theta_1 = 2\lambda_3-1\,,
 \end{align}
 with
 \begin{align}
\rho&=\frac{m_c^2}{m_b^2}, &\kappa &= 1- \left(1- \cos\theta_1\right) y/2 \,.
\end{align}
In terms of these quantities, the phase-space integral in $d=4-2\epsilon$ reads
 \begin{multline}
 \int_{m_c^2}^{m_b^2} \frac{d p_x^2}{2\pi}\int \left [ {\rm d} \Pi_{b\rightarrow x + \ell + \bar\nu}\right] \\
 =\frac{\Omega_{d-1}\Omega_{d-2}\,m_b^{4-4\epsilon}}{2^{d+1}(2\pi)^{2d-2}} \int_0^{1-\rho}\!\!\!\!\!dy\,\int_0^1 d\lambda_2 d\lambda_3 (1-\rho-y)^{2-2 \epsilon } \kappa^{2 \epsilon
   -2}   \left(y  (1-\lambda_2)\right)^{1-2 \epsilon } \left((1-\lambda _3) \lambda_3\right)^{-\epsilon }\,, \label{eq:leptonicphase}
\end{multline}
 with the $d$-dimensional solid angle
\begin{equation}
\Omega_d=\frac{2\pi^{d/2}}{\Gamma(d/2)}\, .
\end{equation}
In the presence of a  cut on the lepton energy $y>y_0$, the transformation $y=(1-\rho-y_0)\lambda_1+y_0$ maps the integration to the unit cube. It is simple to obtain the tree-level phase space from (\ref{eq:leptonicphase}). To this end, one multiplies with $2\pi \delta(p_x^2-m_c^2)=2\pi \delta\left((1-\rho-y)\lambda_2\right)$ and integrates over $\lambda_2$. 

\subsection{Two-body phase space $x\rightarrow c+g$}

We split the gluon three-momentum into a part in the direction of $\vec{p}_x$ and an orthogonal part
\begin{equation}
p_g=(E_g,0,0,0)+ E_g \cos\theta_2\, (0,\frac{\vec{p}_x}{|\vec{p}_x |})+ E_g \sin\theta_2\, (0,\vec{p}_\perp)\,
\end{equation}
 with $\vec{p}_\perp\cdot \vec{p}_x=0$ and $\vec{p}^2_\perp=1$. Expressed in terms of these quantities, the two-body phase space is
\begin{equation}
\int \left [ {\rm d} \Pi_{x\rightarrow c+g}\right] =\frac{1}{2(2\pi)^{d-2}}\int\! 
d\cos\theta_2\, \frac{\sin^{d-4}\!\theta_2\, E_g^{d-2}}{p_x^2-m_c^2}\,\int d^{d-2}p_\perp\,.\label{eq:hadronicphase}
\end{equation}
For a given angle $\theta_2$, the gluon energy is
\begin{equation}
E_g=\frac{p_x^2-m_c^2}{2(E_x- \cos\theta_2 |\vec{p}_x|)}\,.
\end{equation}
Note that the denominators of the real-emission diagrams
\begin{align}\label{props}
(p_b-p_g)^2-m_b^2& =-2 m_b E_g \,, \\
(p_c+p_g)^2-m_c^2&=p_x^2-m_c^2
\end{align}
are independent of $p_\perp$. The only dependence on $p_\perp$ arises from the scalar products with lepton momenta in the numerator of the diagrams. The integration over the unit vector $p_\perp$ is therefore trivial. The only non-vanishing integrals we need are
\begin{equation}
\int d^{d-2}p_\perp 
\Bigg \{1, p_\perp^i p_\perp^j \Bigg\} = 
\Bigg\{ 1 , \frac{1}{d-2}\, \delta^{ij} \Bigg\} \Omega_{d-2}\,,
\end{equation}
where $\delta^{ij}$ is the metric on the $(d-2)$-dimensional sub-space, with $\delta^{i}_{i}=d-2$. For the evaluation of the diagrams it is simplest to parameterize the vector $p_\perp$ as
\begin{equation}
p_\perp=\sin\theta_3\,(0,0,1,0)+\cos\theta_3 \frac{1}{|\vec{p}_x|}(0,E_\nu \cos\theta_1+E_l,0,-E_\nu \sin \theta_1)\, .
\end{equation}
The integrand is then a second-order polynomial in $\cos\theta_3$ and $\sin\theta_3$ and the integral over $p_\perp$ takes the form
\begin{equation} \label{eq:perp}
\int d^{d-2}p_\perp\,\Bigg\{ 1, \cos^2\theta_3 \Bigg\} = \Omega_{d-3} \int_{-1}^1\!d\cos\theta_3\, \sin^{d-5}\!\theta_3 \,\Bigg\{ 1,\cos^2\theta_3\Bigg\} = \Omega_{d-2} \,\Bigg\{ 1, \frac{1}{d-2} \Bigg\}\, .
\end{equation}
To calculate the rate, we combine
(\ref{eq:leptonicphase}) with (\ref{eq:hadronicphase}) and (\ref{eq:perp}) and rewrite $\cos\theta_2=2\lambda_4-1$.

The point $\lambda_2=0$ corresponds to the kinematic configuration 
where soft singularities occur, since $E_g \propto
p_x^2-m_c^2=(1-\rho-y)\lambda_2 \rightarrow 0$. Both propagator denominators (\ref{props}) are proportional to $\lambda_2$ and vanish at this point. The phase space (\ref{eq:hadronicphase}) itself is proportional to $\lambda_2^{1-2\epsilon}$ so that the infrared divergences take the form
\begin{equation}
\frac{1}{\lambda_2^{1+2\epsilon}}=-\frac{1}{2\epsilon}\delta(\lambda_2)+\left[\frac{1}{\lambda_2}\right]_+ +{\cal O}(\epsilon)\, .
\end{equation}
The above relation is easily implemented into the code for the numerical evaluation of the diagrams. We evaluate both the divergent and the finite part numerically and check that the $1/\epsilon$ divergences cancel in the final result within our numerical accuracy.

\subsection{Loop integrals}
 
 The virtual corrections involve loop integrals 
\begin{equation}
\left\{I,I_\mu,I_{\mu\nu}\right\}= \int d^d k \frac{\left\{1,k_\mu,k_\mu k_\nu\right\} }{k^2\,(2p_b \cdot  k+k^2)\,(2p_c \cdot  k+k^2+p_c^2-m_c^2)}
 \end{equation}
 with $p_b^2=m_b^2$. We need the loop integrals for $p_c^2\neq m_c^2$ because we replace $p_c^\mu\rightarrow p_c^\mu +r^\mu$ and then expand in the residual momentum $r_\mu$. The Feynman parameterization of the integral has the form
 \begin{multline}\label{feynp}
 \left\{I,I^\mu,I^{\mu\nu}\right\} =i\pi^{d/2}\Gamma(1+\epsilon) \int_0^1\!\! du\,dv \, v\, \Delta^{-1-\epsilon}\\ \left\{-1, v\,(u p_b^\mu+\bar{u} p_c^\mu), -v^2\,(u p_b^\mu+\bar{u} p_c^\mu)(u p_b^\nu+\bar{u} p_c^\nu)+\frac{1}{2\epsilon}\Delta\, g^{\mu\nu} \right\}
\end{multline}
with $\bar u= 1-u$ and
 \begin{equation}
 \Delta= v^2\, \left [m_b^2 u^2+ m_c^2 \bar{u}^2+  2 p_b\cdot p_c \bar{u} u\right] + v\,\bar u \, \left(1+\bar{u} v\right) \left(m_c^2-p_c^2\right)\,.
\end{equation}
Since we expand around the mass shell, the integral over the Feynman parameter $v$ can always be done analytically after expanding, because the $v$ integration completely factors after setting $p_c^2=m_c^2$. For the scalar integral, the $v$ integration produces $1/\epsilon$ infrared divergences. The $u$ integration on the other hand is always finite and done numerically, together with the integration over the tree-level phase space.

\begin{table}[t]
\begin{center}
\begin{tabular}{cccccc}\toprule
 & $1$ & $\frac{\alpha_s}{\pi}$ & $\frac{\mu_\pi^2}{2m_b^2}$ 
 &  $\frac{\alpha_s}{\pi}\,\frac{\mu_\pi^2}{2m_b^2}$ & \%  \\ \midrule
\gr $1$ & $0 .6319 (4) $ & $ - 1.123 (4) $ & $ - 
     0.6319 (6) $ & $1 .125 (8) $ &  $0 .1 $   \\
 $ {\hat E_l} $ & $0 .1941 (1) $ & $ - 0.348 (1) $ & $0 .0000 (3) $ & $0 .000 (3) $ &  $0.$   \\ \gr
 $ {\hat E} _l^2  $ & $0 .06509 (5) $ & $ - 0.1186 (5) $ & $0 .1085 (1) $ & $ - 0.198 (1) $ &  $ - 
 0.2 $  \\
 $  {\hat E} _l^3  $ & $0 .02308 (2) $ & $ - 
         0.0429 (2) $ & $0 .09232 (5) $ & $ - 0.1714 (7) $ &  $ - 
 0.5 $  \\ \gr
 $  {\hat E} _x $ & $0 .2667 (2) $ & $ - 0.454 (2) $ & $ - 
     0.6319 (2) $ & $1 .124 (3) $ &  $0 .3 $  \\
 $  {\hat E} _x^2  $ & $0 .11576 (9) $ & $ - 0.1845 (9) $ & $ - 
     0.3667 (1) $ & $0 .610 (2) $ &  $0 .4 $  \\ \gr
 $  {\hat E} _x^3   $ & $0 .05148 (4) $ & $ - 0.0744 (4) $ & $ - 
     0.17468 (6) $ & $0 .2534 (8) $ &  $0 .4 $  \\
 $  ({\hat p} _x^2 - \rho)  $ & $0$ & $0 .05693 (3) $ & $ - 
      0.7305 (2) $ & $1 .281 (3) $ &  $ - 41. $  \\ \gr
 $  ({\hat p} _x^2 - \rho)^2  $ & $0$ & $0 .005754 (3) $ & $0 .20337 (5) $ & $ - 0.5712 (9) $ &  $ - 19.4 $  \\
 $  ({\hat p} _x^2 - \rho)^3  $ & $0$ & $0 .0011438 (6) $ & $0$ & $0.036918(7) $ &  $23 .4 $  \\ \gr
 $  {\hat E} _x ({\hat p} _x^2 - \rho)  $ & $0$ & $0 .02970 (2) $ & $ - 
     0.20013 (6) $ & $0 .2544 (8) $ &  $47 .2 $  \\
 $  {\hat E} _x ({\hat p} _x^2 - \rho)^2  $ & $0$ & $0 .003373 (2) $ & $0 .09285 (2) $ & 
$ - 0.2455 (4) $ &  $ - 17.1 $  \\   \gr 
 $   {\hat E} _x^2 ({\hat p} _x^2 - \rho) $ & $0$ & $0 .015856 (8) $ & $ - 
       0.03570 (2) $ & $ - 0.0208 (3) $ &  $ - 1.8 $   \\ \bottomrule
\end{tabular}
\end{center}
\caption{Coefficients of the perturbative and power corrections to the moments (\ref{momentsdef}) without a cut on the lepton energy for $m_c/m_b=1/4$. Perturbative corrections are given in units of $\alpha_s/\pi$, the power corrections in units of $\mu_\pi^2/(2m_b^2)$. All entries need to be multiplied by the common factor $G_F^2 |V_{cb}|^2 m_b^5 /(192\pi^3)$. The numbers in the table correspond to the partonic moments in the pole scheme. The last column gives the relative size of the kinetic ${\cal O}(\alpha_s)$ corrections for default values of the parameters, see text.  \label{tab:res00}}
\end{table}

\begin{table}[th]
\begin{center}
\begin{tabular}{cccccc}\toprule
& $1$ & $\frac{\alpha_s}{\pi}$ & $\frac{\mu_\pi^2}{2m_b^2}$ 
 &  $\frac{\alpha_s}{\pi}\,\frac{\mu_\pi^2}{2m_b^2}$ & \%  \\ \midrule
 \gr $1$ & $0 .5149 (3) $ & $ - 0.910 (3) $ & $ - 
     0.5692 (6) $ & $0 .987 (8) $ &  $0 .1 $  \\
 $ {\hat E_l} $ & $0 .1754 (1) $ & $ - 
        0.314 (1) $ & $0 .0109 (3) $ & $ - 0.024 (3) $ &  $0.$  \\  \gr
 $ {\hat E} _l^2 $ & $0 .06189 (5) $ & $ - 
         0.1128 (5) $ & $0 .1105 (1) $ & $ - 0.202 (1) $ &  $ - 
 0.2 $  \\
 $ {\hat E} _l^3 $ & $0 .02251 (2) $ & $ - 
         0.0418 (2) $ & $0 .09269 (5) $ & $ - 0.1722 (7) $ &  $ - 
 0.6 $  \\  \gr
 $ {\hat E} _x$ & $0 .2111 (1) $ & $ - 0.365 (1) $ & $ - 
     0.5694 (2) $ & $1 .010 (3) $ &  $0 .4 $  \\
 $ {\hat E} _x^2 $ & $0 .08917 (7) $ & $ - 0.1482 (7) $ & $ - 
     0.3378 (1) $ & $0 .576 (1) $ &  $0 .5 $  \\  \gr
 $ {\hat E} _x^3 $ & $0 .03867 (4) $ & $ - 0.0606 (4) $ & $ - 
     0.16898 (6) $ & $0 .2639 (7) $ &  $0 .5 $  \\
 $ ({\hat p} _x^2 - \rho) $ & $0$ & $0 .03618 (2) $ & $ - 
      0.6855 (2) $ & $1 .213 (2) $ &  $ - 25.5 $  \\  \gr
 $ ({\hat p} _x^2 - \rho)^2 $ & $0$ & $0 .002808 (2) $ & $0 .15198 (4) $ & $ - 0.4388 (5) $ &  $ - 21.6 $  \\
 $ ({\hat p} _x^2 - \rho)^3 $ & $0$ & $0 .0004053 (3) $ & $0$ & $0 .020998 (4) $ &  $32 .9 $  \\  \gr
 $ {\hat E} _x ({\hat p} _x^2 - \rho) $ & $0$ & $0 .01801 (1) $ & $ - 
      0.20707 (6) $ & $0 .2961 (8) $ &  $ - 39.2 $  \\
 $ {\hat E} _x ({\hat p} _x^2 - \rho)^2 $ & $0$ & $0 .0015307 (10) $ & $0 .06794 (2) $ 
& $ - 0.1897 (3) $ &  $ - 20.1 $  \\  \gr
 $ {\hat E} _x^2 ({\hat p} _x^2 - \rho) $ & $0$ & $0 .009147 (6) $ & $ - 
     0.05271 (2) $ & $0 .0304 (3) $ &  $12 .4 $ \\ \bottomrule
\end{tabular}
\end{center}
\caption{Coefficients of the perturbative and power corrections to the the moments (\ref{momentsdef}) with $4.6 {\hat E_l}>1$ and $m_c/m_b=1/4$. Perturbative corrections are given in units of $\alpha_s/\pi$, the power corrections in units of $\mu_\pi^2/(2m_b^2)$. All entries need to be multiplied by the common factor $G_F^2 |V_{cb}|^2 m_b^5 /(192\pi^3)$. The numbers in the table correspond to the partonic moments in the pole scheme. The last column gives the relative size of the kinetic ${\cal O}(\alpha_s)$ corrections for default values of the parameters, see text.  \label{tab:res04}}
\end{table}

\begin{table}[th]
\begin{center}
\begin{tabular}{ccccccc}\toprule
& $1$ & $v$ & $w$  &   $v^2$ & $w^2$ & $v \; w$ 
  \\ \midrule \gr
 $1$ & $0.860$ & $-1.385$ & $0.453$ & $-1.216$ & $-1.463$ &  $-0.064$  \\
 ${\hat E_l}$ & $-0.056$ & $-0.007$ & $0.025$ & $-0.342$ & $-0.478$ &  
$-0.012$  \\ \gr

 ${\hat E}_l^2$ & $-0.210$ & $0.335$ & $-0.170$ & $-0.098$ & $-0.153$ &  
$-0.002$  \\
 ${\hat E}_l^3$ & $-0.174$ & $0.331$ & $-0.210$ & $-0.029$ & $-0.048$ &  $0.000$  
\\ \gr

 $ {\hat E}_X$ & $0.938$ & $-1.356$ & $0.413$ & $-0.640$ & $-0.624$ &  
$-0.011$  \\
 $ {\hat E}_X^2$ & $0.553$ & $-0.598$ & $0.006$ & $-0.212$ & $-0.221$ &  
$-0.172$  \\ \gr
 $ {\hat E}_X^3$ & $0.264$ & $-0.172$ & $-0.095$ & $-0.013$ & $-0.058$ &  
$-0.192$  \\
 $ ({\hat p}_X^2-\rho)$ & $1.191$ & $-1.787$ & $0.706$ & $-0.182$ & $-0.086$ 
&  $0.009$  \\ \gr
 $ ({\hat p}_X^2-\rho)^2$ & $-0.39$ & $0.745$ & $-0.476$ & $0.392$ & $0.188$ 
&  $-0.449$  \\
 $ ({\hat p}_X^2-\rho)^3$ & $0.017$ & $-0.046$ & $0.05$ & $-0.033$ & $-0.001$ 
&  $0.068$  \\ \gr
 ${\hat E}_X ({\hat p}_X^2-\rho)$ & $0.314$ & $-0.25$ & $-0.109$ & $0.136$ & 
$0.063$ &  $-0.267$  \\
 ${\hat E}_X ({\hat p}_X^2-\rho)^2$ & $-0.169$ & $0.278$ & $-0.126$ & $0.165$ 
& $0.081$ &  $-0.146$  \\ \gr
 ${\hat E}_X^2 ({\hat p}_X^2-\rho)$ & $0.050$ & $0.067$ & $-0.125$ & $0.153$ & 
$0.068$ &  $-0.195$  \\ \gr 
 \bottomrule
\end{tabular}
\end{center}
\caption{
Dependence of the coefficient of $\frac{\alpha_s}{\pi} \frac{\mu_\pi^2}{2m_b^2}$ of the moments (\ref{momentsdef}) on the lepton-energy cut and the charm-quark mass. We define $v \equiv 4 m_c/m_b -1$ and $w \equiv (4E_0 - m_b)/m_b$ and expand the moments to second order in these variables. The expansion coefficients in the table were determined by performing a quadratic fit to the exact results in the range $0.2 \leq m_c/m_b \leq 0.3$ and $0.5 \leq 4.6\,{\hat E}_0 \leq 1.6$. \label{interpolation}}
\end{table}

\section{Results for the moments of the differential rate\label{results}}

When doing the calculation, it is simplest to evaluate moments using partonic variables. To distinguish partonic and hadronic quantities, we denote the partonic energy and invariant mass by $E_x$ and $p_x^2$, while writing $E_X$ and $p_X^2$ in the hadronic case. For the tree-level diagrams $E_x$ is the energy of the charm quark, while in the diagrams where a gluon is emitted $E_x=E_c+E_g$. The partonic moments for which we present results in tables 
are defined as
\begin{equation}\label{momentsdef}
\left[ w(E_l, E_x, p_x^2) \right] = \int_{E_0}^{E_{\rm max}}\!\!\! dE_l \int\!\! dE_x\, dp_x^2 \,\frac{d\Gamma}{dE_x\, dp_x^2\, dE_l } w(E_l, E_x, p_x^2)\,.
\end{equation}
We consider lepton energy moments $w={\hat E}_\ell^n=(E_\ell/m_b)^n$ with $n=1\dots 3$ and the partonic energy and invariant mass moments $w={\hat E}_x^n\,({\hat p}_x^2-\rho)^m$ with $n+m\leq 3$ and $\rho=m_c^2/m_b^2$. Note that we do not normalize the partonic moments to the rate. 

Numerical results for the moments without a cut and with a cut $\hat E_\ell>1/4.6$ (corresponding to $E_\ell>1.0\,{\rm GeV}$ for $m_b=4.6 \,{\rm GeV}$) are shown in tables \ref{tab:res00} and \ref{tab:res04} for $\sqrt{\rho}=m_c/m_b=1/4$. In the last column, we indicate the relative size of the kinetic ${\cal O}(\alpha_s)$ corrections. To estimate the relative size we use $\alpha_s=0.22$ and $\mu_\pi^2=0.4 \,{\rm GeV}^2$. As explained in Section \ref{kinetic}, we calculate the moments with two different methods and check that the results agree within numerical precision. In the tables, we also include numerical results  for the tree-level moments. Using (\ref{eq:momrel}) it would be simple to evaluate them analytically.

The one-loop kinetic corrections are small for the moments which get contributions at leading power but sizable for the moments of $({\hat p}_x^2-\rho)^n$. For example, the moment of $p_x^2-\rho$, gets a correction of $-40\%$. Compared to the tree-level contributions of the kinetic operator, the one-loop terms are typically suppressed by a few times $\alpha_s/\pi$. We thus expect that the extracted value of $\mu_\pi^2$ will be shifted by about $\pm20\%$ by their presence. Given the size of the kinetic corrections to the rate, we do not expect that the ${\cal O}(\alpha_s)$ corrections will affect the extracted value of $|V_{cb}|$. In the fit of Ref.~\cite{Buchmuller:2005zv} the value of $\mu_\pi^2$ is varied by $\pm 20\%$ to estimate the theoretical uncertainty. The corrections we calculate are indeed of this size, except that varying the value of $\mu_\pi^2$  correlates the change in all moments, while the perturbative corrections are different in each case. In the fit of \cite{Bauer:2004ve}, the corrections are underestimated to be $\frac{\alpha_s}{4\pi}\Lambda_{\rm QCD}^2/m_b^2\sim 0.0002$: the contributions we find are roughly ten times larger.

In Table \ref{interpolation} we give the result for the one-loop kinetic corrections as a function of the cut energy $E_0$ and the ratio $\rho=m_c^2/m_b^2$. To this end, we perform a quadratic fit around default values $m_c/m_b=1/4$ and $E_0=m_b/4\approx 1.15\, {\rm GeV}$. The accuracy of the quadratic fit is a few per cent except in cases where the corrections become very small. Tables with precise numerical results for arbitrary cut energies and charm-mass values can be obtained from the authors.

Instead of the partonic moments, experimental papers present results for the normalized hadronic moments
\begin{equation}
\left\langle w(E_l, E_X, p_X^2) \right\rangle = \frac{1}{\Gamma(E_l>E_0)} \int_{E_0}^{E_{\rm max}}\!\!\! dE_l \int\!\! dE_X\, dp_X^2 \,\frac{d\Gamma}{dE_X\, dp_X^2\, dE_l}\, w(E_l, E_X, p_X^2)\,.
\end{equation}
To translate the results to hadronic kinematics we note that leptonic quantities are identical on the hadronic and partonic level. Using that the $B$-meson momentum is $p_B^\mu=M_B\,v^\mu$, it follows that
\begin{align}
E_X &= M_B - v\cdot q = M_B-m_b + E_x  \,,\\
p_X^2 & = (p_B-q)^2 
= p_x^2 + 2 E_x (M_B-m_b)+(M_B-m_b)^2\,.
\end{align}
With these two equations, it is straightforward to translate our partonic results into hadronic language. For example, the prediction for the lowest moments are obtained from the relations
\begin{align}
\langle E_\ell \rangle &=  \frac{1}{\left[ 1 \right]} m_b \left[ \hat {E}_\ell \right] \,, \nonumber\\
\langle E_X \rangle &=  \frac{1}{\left[ 1 \right]} 
\left( m_b \left[{\hat E}_x \right] + (M_B-m_b)\,\left[ 1 \right]\right)\,,\\
\langle p_X^2-M_D^2 \rangle &=  \frac{1}{\left[ 1 \right]} 
\left( m_b^2\,\left[\hat{p}_x^2-\rho\right] + 2 m_b (M_B-m_b)\,\left[{\hat E}_x \right] + ((M_B-m_b)^2+(m_c^2-M_D^2))\,\left[ 1 \right]\right)\,. \nonumber
\end{align}
The moment with unit weight function is the rate: $\left[1\right]=\Gamma(E_\ell>E_0)$, see (\ref{momentsdef}).
While the moments of $p_x^2-m_c^2$ vanish at tree level in the heavy-quark limit, the hadronic moments $p_X^2-M_D^2$ are nonzero. Using the above conversion formulae together with the results in Table \ref{tab:res04} we obtain for example
\begin{align}
\langle p_X^2-M_D^2 \rangle &= \left[ 0.860 + 1.59 \frac{\alpha_s}{\pi}+ \left(-32.3 +1.96 \frac{\alpha_s }{\pi} \right)\frac{\mu_\pi^2}{2m_b^2}\right] \,{\rm GeV}^2\,, \\
\langle (p_X^2-M_D^2)^2 \rangle &= \left[ 0.939 + 7.00 \frac{\alpha_s}{\pi}+\left(117.4 -178.2 \frac{\alpha_s }{\pi}\right)\frac{\mu_\pi^2}{2m_b^2}\right] \,{\rm GeV}^4 \,.
\end{align}

Finally, let us note that we performed our calculation in the pole scheme, but it is simple to convert our result into different schemes. The pole scheme is calculationally most convenient, but plagued by large higher-order corrections. The problem arises because the definition of the parameters in this scheme relies on on-shell quark states, a concept not meaningful beyond perturbation theory. The resulting bad perturbative behavior can be improved by using parameter definitions with less infrared sensitivity such as $\overline{\rm MS}$ quark masses. This is appropriate for the charm quark which we treat as light, however, the $\overline{\rm MS}$ mass definition is not suitable for the $b$-quark, because it is not consistent with HQET power counting. A number of alternative schemes,  appropriate for heavy-quark processes, are available: they include the kinetic \cite{Uraltsev:1996rd}, the potential-subtracted \cite{Beneke:1998rk}, the $1S$ \cite{Hoang:1998hm} and the shape-function scheme \cite{Bosch:2004th}. To one-loop order, the scheme changes from the pole into the new schemes have the form
\begin{align}
m_b &= m_b(\mu_f) + \mu_f \frac{\alpha_s}{\pi} c_1 +  \frac{\mu_f^2}{2m_b^2} \frac{\alpha_s}{\pi} c_2  \,, &
\mu_\pi^2 &= \mu_\pi^2(\mu_f)\left(1+\frac{\alpha_s}{\pi}\,c_3\right) +\mu_f^2 \frac{\alpha_s}{\pi} c_4  \,,
\end{align}
where coefficients $c_i$ depend on the scheme. For example, in the kinetic scheme $c_1=\frac{4}{3}\,C_F$, $c_2=C_F$, $c_3=0$, $c_4=C_F$ and $\mu_f=1\,{\rm GeV}$. The exact choice of the factorization scale $\mu_f$ is a matter of convention. When performing the moment fit, it is worthwhile to check to what extent the fit results are independent of this choice. A compendium of two-loop scheme conversion formulae can be found in \cite{Neubert:2004sp}. Let us note that only the kinetic and the shape-function scheme provide improved definitions for the parameter $\mu_\pi^2$. To obtain the scheme change to ${\cal O}(\alpha_s)$ one simply replaces the pole-scheme parameters by the redefined ones in our one-loop results. Additional ${\cal O}(\alpha_s)$ corrections are generated when performing the scheme change in the tree-level results. We refrain here from explicitly changing the scheme, but the default values of our parameters are chosen such that they correspond to values which are typical for the improved schemes.

\section{Summary and conclusion}

We have evaluated the one-loop perturbative corrections to the coefficient of the kinetic operator in the operator product expansion of the decay $\bar B\rightarrow X_c\ell \bar \nu$. The corrections are typically $(1-3)\times \frac{\alpha_s}{\pi}\approx 10-30\%$ times the leading kinetic power correction. We thus expect that these corrections will change the extracted value of $\mu_\pi^2$ from the moment fit by about $20\%$. Whether this in turn has an effect on the extracted $m_b$ and $m_c$ values is hard to estimate without performing the global fit. Since the kinetic corrections are very small for the total rate, the value of $|V_{cb}|$ will likely not be affected.

With the same numerical methods used here the one-loop chromo-magnetic and the two-loop leading-power corrections can be calculated as well. Once these are known, the theoretical precision on the predictions for the $\bar B\rightarrow X_c\ell\bar\nu$ decay will be superior to the experimental accuracy. These results will increase the precision of the extracted parameters and provide a nontrivial consistency check on the experimental data used in the fit. The calculation will also answer the question whether the currently used theoretical error estimates on the extracted parameters are realistic. This is of particular importance because the value and uncertainty of $m_b$ is a crucial input in the determination of $|V_{ub}|$ from the inclusive $\bar B\rightarrow X_u\ell \bar\nu$ decay. 

\subsubsection*{Acknowledgments}

We thank R.~Feger, T.~Feldmann, T.~Mannel, K.~Melnikov, M.~Neubert and G.~Paz for discussions. H.~B.~thanks the Fermilab Theoretical Physics Department for hospitality. Fermilab is operated by Fermi Research Alliance, LLC under Contract No. DE-AC02-07CH11359 with the United States Department of Energy. 

\begin{appendix}

\section{Moment relations \label{app:momrel}}

Without cuts on the available phase space, the kinetic corrections to the moments are directly related to the leading-power moments. The relations can be derived from the general result (\ref{eq:momrel}). For convenience, we list here the explicit form of the relations for the moments we are interested in. We write the relations as $A \equiv B $ which should be read as
\begin{equation}
\int\! {\rm d}E_l  {\rm d}E_x\, {\rm d}p_x^2 \,\frac{d\Gamma}{dE_x\, dp_x^2\, dE_l }\, A = \int {\rm d}E_l  {\rm d}E_x\, {\rm d}p_x^2 \,\frac{{\rm d}\Gamma^{\rm partonic}}{dE_x\, dp_x^2\, dE_l } \, B \,.
\end{equation}
For the energy moments the relations read
{\allowdisplaybreaks
\begin{align}
 E_\ell & \equiv E_\ell  \,,&
 E_\ell^2 & \equiv \left(1+\frac{5}{3} \frac{\mu_\pi^2}{2m_b^2}\right)  
 E_\ell^2  \,, \nonumber \\
 E_\ell^3 & \equiv \left(1+ 4 \frac{\mu_\pi^2}{2m_b^2}\right) E_\ell^3 \,, &
  E_x & \equiv E_x-\frac{\mu_\pi^2}{2m_b^2}  \,, \\
 E_x^2 & \equiv E_x^2+\frac{\mu_\pi^2}{ 2m_b^2} \left(-\frac{2}{3} p_x^2+\frac{5}{3} E_x^2-2 E_x m_b\right) \,,&
 E_x^3 & \equiv E_x^3+\frac{\mu_\pi^2}{2m_b^2} \left(4 E_x^3-3 E_x^2 m_b-2  p_x^2 E_x\right) \,, \nonumber
 \end{align}
 and for the partonic invariant mass moments the relations are
 \begin{align}
 p_x^2 & \equiv p_x^2+\frac{\mu_\pi^2}{2m_b^2} \left(-p_x^2+2 E_x m_b-2 m_b^2\right) \,, \nonumber \\
  (p_x^2)^2 & \equiv  (p_x^2)^2+\frac{\mu_\pi^2}{2m_b^2} \left(- (p_x^2)^2+4 E_x p_x^2 m_b-\frac{ 20}{3} p_x^2 m_b^2+\frac{8 }{3}E_x^2 m_b^2\right) \,, \nonumber \\
  (p_x^2)^3 & \equiv 
       (p_x^2)^3+\frac{\mu_\pi^2}{2m_b^2} \left(- (p_x^2)^3+6  E_x  (p_x^2)^2 m_b-14  (p_x^2)^2 m_b^2+8 E_x^2 p_x^2 m_b^2\right)  \,,\nonumber \\
 E_x  p_x^2& \equiv E_x p_x^2+\frac{\mu_\pi^2}{2m_b^2} \left(-\frac{7 }{3}p_x^2 m_b+\frac{10 }{3}E_x^2 m_b-2 E_x m_b^2\right) \,,\\
   E_x (p_x^2)^2 & \equiv E_x   (p_x^2)^2+\frac{\mu_\pi^2}{2m_b^2} \left(-\frac{11}{3} (p_x^2)^2 m_b+\frac{20}{3} E_x^2 p_x^2 m_b-\frac{20}{3}   E_x p_x^2 m_b^2+\frac{8 }{3}E_x^3 m_b^2\right)  \nonumber \,, \\
 E_x^2  p_x^2 & \equiv E_x^2 p_x^2 +\frac{\mu_\pi^2}{2m_b^2} \left(-\frac{2 }{3} (p_x^2)^2+\frac{5}{3} E_x^2 p_x^2-\frac{ 14}{3} E_x p_x^2 m_b+\frac{14}{3} E_x^3 m_b-2 E_x^2 m_b^2\right) \, . \nonumber
\end{align}}

\end{appendix}

\end{document}